\documentclass[a4paper,11pt]{article}
\usepackage{pos}

\usepackage{subcaption}
\usepackage{wrapfig}
\usepackage{microtype}

\title{Dimuon production in neutrino-nucleus collisions - the SIDIS approach}

\author{I. Helenius}
\author{H. Paukkunen}
\author*{S. Yrjänheikki}

\affiliation{University of Jyväskylä, Department of Physics, P.O. Box 35, 40014 University of Jyväskylä, Finland}

\affiliation{Helsinki Institute of Physics, P.O. Box 64, 00014 University of Helsinki, Finland}

\emailAdd{ilkka.m.helenius@jyu.fi}
\emailAdd{hannu.t.paukkunen@jyu.fi}
\emailAdd{sami.a.yrjanheikki@jyu.fi}

\abstract{Dimuon production is in many global parton distribution function analyses calculated by assuming that it is proportional to inclusive charm production. As this assumption breaks down at next-to-leading order in the perturbative expansion, we present a direct calculation of dimuon production that does not require an external acceptance correction. Our calculation utilizes semi-inclusive deep inelastic scattering and a decay function fitted to experimental data. We find our calculation to be in good agreement with available experimental data. Here we also demonstrate that the acceptance correction depends on the used parton distribution and perturbative order.}

\FullConference{31st International Workshop on Deep Inelastic Scattering (DIS2024)\\
 8–12 April 2024\\
Grenoble, France\\}

\renewcommand{\hookAfterAbstract}{%
	\par\bigskip
	\textsc{ArXiv ePrint}:
	\href{https://arxiv.org/abs/2405.12677}{2405.12677}
}

\newcommand{\msbar}{\overline{\mathrm{MS}}}

\newcommand{\dd}{\mathrm{d}}
\newcommand{\Gammatot}{\Gamma_{\mathrm{tot}}}

\begin{document}
\maketitle

\setlength{\abovedisplayskip}{7pt}
\setlength{\belowdisplayskip}{7pt}

\section{Introduction}

In global analyses of parton distribution functions (PDFs), the strange-quark distribution remains poorly constrained \cite{Hou:2019efy,Bailey:2020ooq,NNPDF:2021njg,Eskola:2021nhw,Duwentaster:2022kpv,AbdulKhalek:2022fyi}. One important process for constraining it is charm production in charged-current deep inelastic scattering (DIS) \cite{CCFR:1994ikl,NuTeV:2001dfo,NuTeV:2007uwm,CHORUS:2008vjb,NOMAD:2013hbk}. This process is typically measured via dimuon production, wherein the charm quark hadronizes into a charmed hadron and subsequently decays into a muon. Dimuon production is typically computed by assuming that it factorizes into inclusive charm production, such that one simply multiplies the charm production cross section with multiplicative correction factors to account for the experimental cuts. While this approach works at the leading order (LO), it breaks down at next-to-leading order (NLO) in the strong coupling constant $\alpha_s$. In our work \cite{Helenius:2024fow} we take an alternative approach and compute dimuon production directly without invoking the aforementioned assumption, but instead use semi-inclusive DIS (SIDIS) for the production of charmed hadrons and introduce a decay function to implement their decay. 

\section{Production of charmed hadrons in neutrino-SIDIS}
\label{sec:sidis}

The neutrino-SIDIS cross section $\nu_\mu(k)+N(P_N)\to \mu(k')+h(P_h)+X$ is obtained by
\begin{equation}
\label{eq:sidis_hadron_cross_section}
\begin{split}
	\frac{\dd\sigma(\nu_\mu N\to \mu h X)}{\dd x \, \dd y \, \dd z}= \frac{G_F^2M_W^4}{\left(Q^2+M_W^2\right)^2}
  \frac{Q^2}{2\pi xy}\bigg[&xy^2 F_1(x, z, Q^2)+\left(1-y-\frac{xy M^2}{s-M^2}\right)F_2(x, z, Q^2) \\ & \pm xy\left(1-\frac{y}{2}\right)F_3(x, z, Q^2)\bigg],
\end{split}
\end{equation}
where the kinematical variables are defined as usual by
\begin{equation}
\label{eq:sidis_kinematics}
\begin{aligned}
	Q^2&=-q^2 = -(k-k')^2, \quad x=\frac{Q^2}{2P_N\cdot q}, \quad y=\frac{P_N\cdot q}{P_N\cdot k}, \quad z=\frac{P_N\cdot P_h}{P_N\cdot q}.
\end{aligned}
\end{equation}
The sign in the $F_3$-term is $+$ for neutrinos and $-$ for antineutrinos. The massless $\msbar$ expressions for the structure functions $F_1$, $F_2$, and $F_3$ can be found in e.g. refs.~\cite{Furmanski:1981cw,deFlorian:1997zj}. In order to take the leading charm-quark mass effects into account, we replace the variable $x$ in the structure functions with the slow-rescaling variable $\chi=x(1+m_c^2/Q^2)$ in channels where a charm quark of mass $m_c$ is produced.

The structure functions are written as double convolutions of hard scattering coefficients, PDFs, and fragmentation functions (FFs). We use the NLO nuclear PDF sets \texttt{EPPS21} \cite{Eskola:2021nhw}, \texttt{nCTEQ15HQ} \cite{Duwentaster:2022kpv}, and \texttt{nNNPDF3.0} \cite{AbdulKhalek:2022fyi}. For the charmed-hadron FFs, we use the NLO sets \texttt{kkks08} \cite{Kneesch:2007ey} for $D^0$ and $D^+$, and \texttt{bkk05} \cite{Kniehl:2006mw} for $D_s$ and $\Lambda_c^+$. In the case of \texttt{kkks08}, we only use the OPAL sets.

\section{Decay of charmed hadrons to muons}

A prominent decay channel for charmed hadrons is to a muon (+ anything), which produces the second muon of the dimuon final state. Assuming that the production of a charmed hadron $h$ and its decay factorizes, we can write the dimuon production cross section as
\begin{equation}
	\sigma(\nu_\mu N\to \mu^-\mu^+ X)=\int \dd\Pi(P_h)\frac{\dd\sigma(\nu_\mu N\to \mu^- hX)}{\dd\Pi(P_h)} \frac{\Gamma_{h\to\mu}}{\Gammatot^h},
\end{equation}
where $\dd\Pi$ is the phase space volume element and $\Gamma_{h\to\mu}$ ($\Gammatot^h$) the partial (total) decay width of the hadron. As part of the decay, we want to include an energy cut on the secondary decay muon, imposed by the experiments \cite{CCFR:1994ikl,NuTeV:2007uwm}. To do this, we define a decay function $d_{h\to \mu}$ such that
\begin{equation}
	\label{eq:decay_width_decay_function}
	\Gamma_{h\to \mu}=\frac{1}{2m_h}\int \frac{\dd^3\mathbf{P}_\mu}{E_\mu} d_{h\to\mu}(w),
\end{equation}
where $w\equiv(P_\mu \cdot P_h)/m_h^2$. Further defining $\rho\equiv E_\mu/E_h$ and neglecting the muon mass, we can use eq.~\eqref{eq:decay_width_decay_function} to define an energy-dependent branching fraction
\begin{equation}
	\label{eq:branching_fraction}
	B_{h\to\mu}(E_h, E_\mu^{\text{min}})\equiv \frac{\Gamma_{h\to\mu}(E_h, E_\mu^{\text{min}})}{\Gammatot^h}\equiv \frac{\pi}{m_h \Gammatot^h}\int \dd\rho \, \rho E_h^2\int \dd(\cos\theta) \, d_{h\to\mu}(w) \, \big|_{E_\mu=\rho E_h\geq E_\mu^{\text{min}}},
\end{equation}
which now includes the experimental energy cut $E_\mu^{\text{min}}$. In the end, the dimuon production cross section can be written as
\begin{equation}
	\frac{\dd\sigma(\nu_\mu N\to \mu^- \mu^+ X)}{\dd x \, \dd y}=\sum_h \int \dd z \, \frac{\dd\sigma(\nu_\mu N\to \mu^- h X)}{\dd x \, \dd y \, \dd z}B_{h\to \mu}(E_h=zyE_\nu, E_{\mu}^{\text{min}}),
\end{equation}
where the sum is over the charmed hadrons $D^0$, $D^+$, $D_s$, and $\Lambda_c^+$. 

In order to obtain the decay function $d_{h\to\mu}$ introduced in eq.~\eqref{eq:decay_width_decay_function}, we parametrize it as
\begin{equation}
	\label{eq:decay_function_parametrization}
	d_{h\to\mu}(w)=Nw^\alpha(1-\gamma w)^\beta \theta(0\leq w\leq 1/\gamma),
\end{equation} 
where $\theta(x)=1$ when $x$ holds, and zero otherwise. From eq.~\eqref{eq:decay_width_decay_function}, we then get
\begin{equation}
	\label{eq:differential_decay_width}
	\frac{\dd\Gamma(h)}{\dd \left|\mathbf P_\mu\right|}=\frac{\pi}{m_h}\left|\mathbf P_\mu\right|\int \dd(\cos\theta) \, d_{h\to\mu}(w).
\end{equation}
The CLEO collaboration has measured this differential decay width distribution in semileptonic D-meson decays $D\to e^+\nu_e X$ for $D^0$ and $D^+$ \cite{CLEO:2006ivk}. While these data are for decays into $e^+$ and not $\mu^+$, 
we neglect any differences between electrons and muons. The decay function is then obtained by fitting eq.~\eqref{eq:decay_function_parametrization} to the CLEO data using eq.~\eqref{eq:differential_decay_width}.

The parameters $N$, $\alpha$, $\beta$, and $\gamma$ in eq.~\eqref{eq:decay_function_parametrization} are highly correlated. To estimate the uncertainty in the resulting decay function, we generate 1000 Monte-Carlo replica data sets from the original CLEO data and perform the same fitting procedure for all replicas to obtain the replica fits. The uncertainty for the dimuon cross section can then be derived by evaluating the cross section for all replica fits. We find that the uncertainty from this fitting procedure is minimal, at around $2\, \%$. 

\section{Results}

\begin{figure}[h!]
	\centering
	\begin{subfigure}{0.3\textwidth}
		\includegraphics[width=\linewidth]{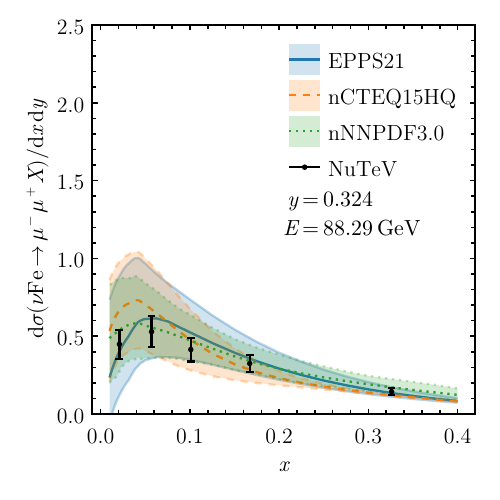}
	\end{subfigure}%
	\begin{subfigure}{0.3\textwidth}
		\includegraphics[width=\linewidth]{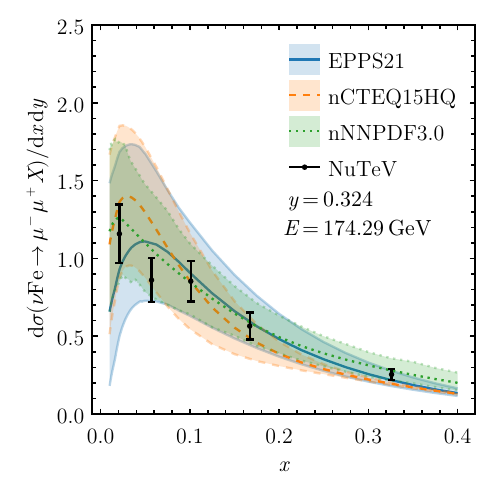}
	\end{subfigure}%
	\begin{subfigure}{0.3\textwidth}
		\includegraphics[width=\linewidth]{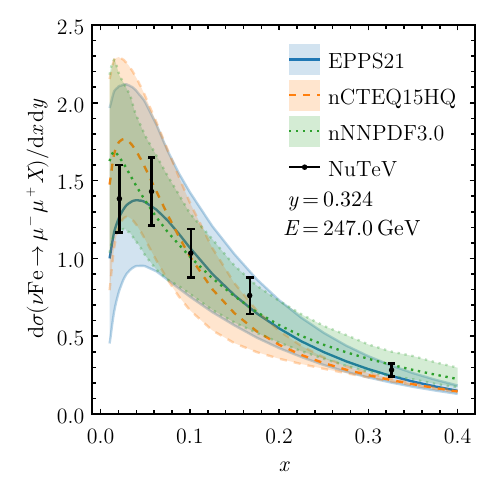}
	\end{subfigure}
	\begin{subfigure}{0.3\textwidth}
		\includegraphics[width=\linewidth]{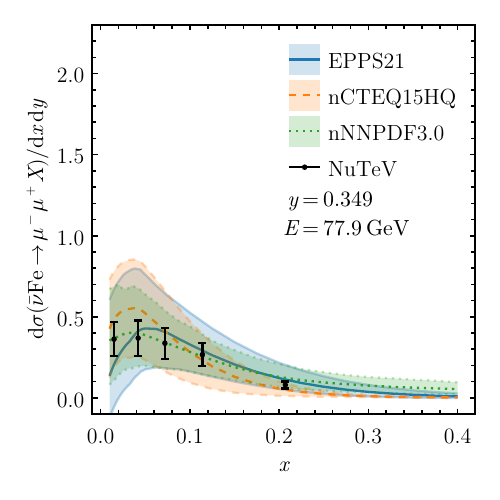}
	\end{subfigure}%
	\begin{subfigure}{0.3\textwidth}
		\includegraphics[width=\linewidth]{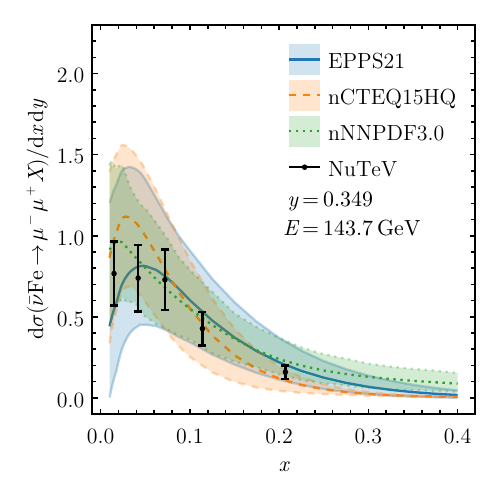}
	\end{subfigure}%
	\begin{subfigure}{0.3\textwidth}
		\includegraphics[width=\linewidth]{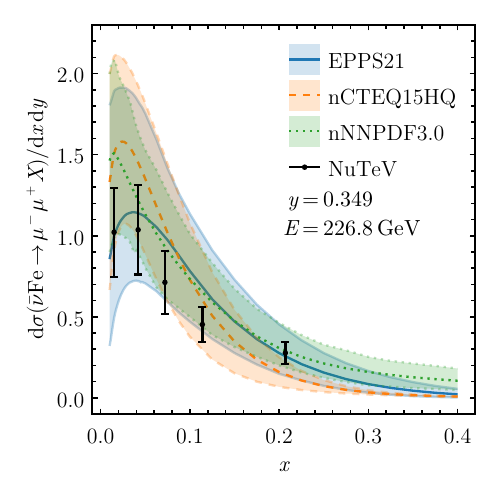}
	\end{subfigure}
	\caption{Neutrino (top row) and antineutrino (bottom row) dimuon cross sections computed at NLO and compared against NuTeV data \cite{NuTeV:2007uwm}. The uncertainty bands indicate the PDF uncertainties with a $90\,\%$ confidence interval. The cross-section values should be multiplied by the factor $G_F^2 ME_\nu/100\pi$.}
	\label{fig:pdf_comparison}
\end{figure}

In our main result, shown in figure~\ref{fig:pdf_comparison}, we compare the dimuon production cross sections with NuTeV data. The calculations are done with three nuclear PDF sets listed in section~\ref{sec:sidis}, allowing for a comparison between the three. Overall, we find good agreement between the calculations and the experimental data. The PDF sets agree within their uncertainty bands, but also have clear differences in their behavior, which reflect the differing $x$-dependence of the $s$-quark distributions among the PDF sets \cite{Klasen:2023uqj}. 

\begin{figure}[h!]
	\centering
	\begin{subfigure}{0.3\textwidth}
		\includegraphics[width=\linewidth]{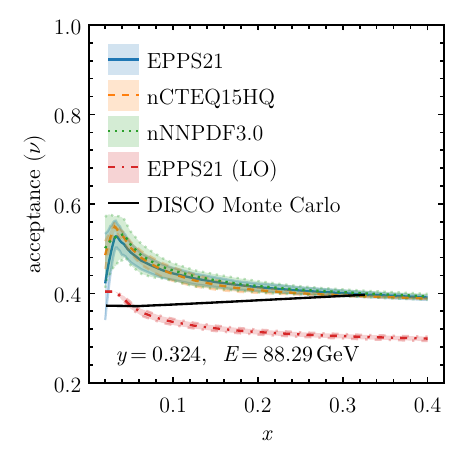}
	\end{subfigure}%
	\begin{subfigure}{0.3\textwidth}
		\includegraphics[width=\linewidth]{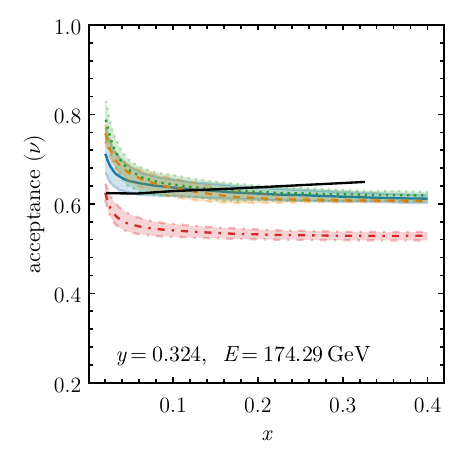}
	\end{subfigure}%
	\begin{subfigure}{0.3\textwidth}
		\includegraphics[width=\linewidth]{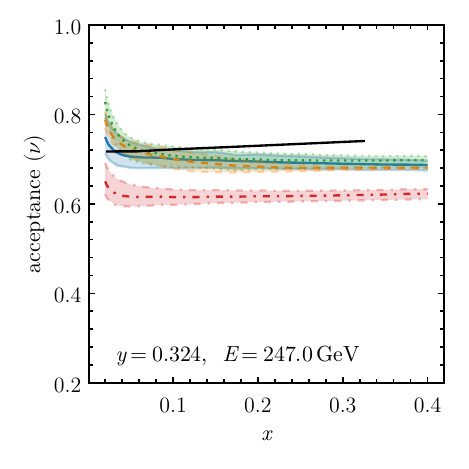}
	\end{subfigure}
	\begin{subfigure}{0.3\textwidth}
		\includegraphics[width=\linewidth]{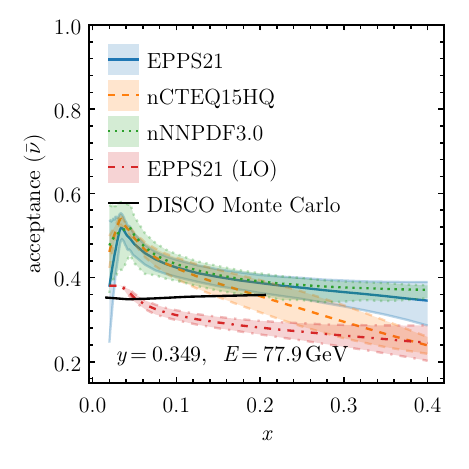}
	\end{subfigure}%
	\begin{subfigure}{0.3\textwidth}
		\includegraphics[width=\linewidth]{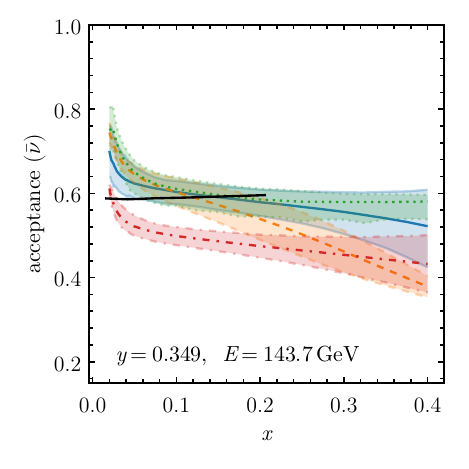}
	\end{subfigure}%
	\begin{subfigure}{0.3\textwidth}
		\includegraphics[width=\linewidth]{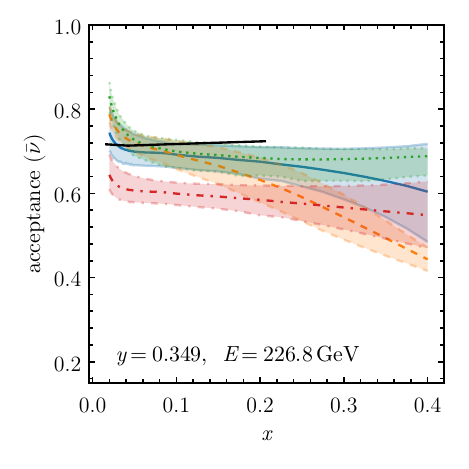}
	\end{subfigure}
	\caption{Effective acceptance correction in neutrino (top row) and antineutrino (bottom row) scattering, computed according to eq.~\eqref{eq:acceptance}, at NLO unless otherwise indicated.	Our calculation is compared against the DISCO Monte-Carlo calculation \cite{Mason:2006qa}.}
	\label{fig:acceptance}
\end{figure}

In calculations of dimuon production based on inclusive charm DIS cross sections, one needs an effective acceptance correction $\mathcal{A}$ as input to take the effects of the muon energy cut into account. In our approach, such a correction is not needed, but can instead be computed as the ratio of dimuon production cross section to inclusive charm production cross section,
\begin{equation}
\label{eq:acceptance}
	\mathcal{A}=\frac{1}{\mathcal{B}_\mu}\frac{\dd\sigma(\nu_\mu N\to \mu\mu X)}{\dd\sigma(\nu_\mu N\to \mu cX)},
\end{equation}
where $\mathcal{B}_\mu=0.092 \pm 10\,\%$ is the average semileptonic branching ratio of charm mesons \cite{Bolton:1997pq}. The acceptance corrections are shown in figure~\ref{fig:acceptance} for (anti)neutrino scattering. Our calculation is also compared against a NLO DISCO Monte-Carlo calculation \cite{Mason:2006qa}. While the two approaches give comparable results, the systematic differences are noticeable. As the neutrino energy increases, our calculation goes from being mostly above the DISCO Monte-Carlo calculation to being mostly below it. The significant differences between the LO and NLO calculations demonstrate the perturbative nature of the effective acceptance. Finally, we note that the acceptance depends on the PDF, as is particularly clear in the antineutrino-scattering case, i.e. the values are tied to a particular PDF.

\section{Summary}

We have presented a SIDIS-based approach for computing dimuon production in neutrino-nucleus collisions. Unlike in the usual approach, we do not need to take an acceptance correction as input to compensate for the muon-energy cut. Thus, the perturbative accuracy can be improved systematically. We demonstrated good agreement between our calculation and available experimental data. We also compared three nuclear PDF sets, which showed general mutual agreement within their uncertainties, but also systematic differences further indicating the sensitivity of dimuon production to the $s$-quark distribution. The effective acceptance correction, which can be directly computed in the presented approach, showed rough agreement with a widely used Monte-Carlo calculation but also clear systematic differences. Furthermore, the acceptance correction was found to be dependent on the used PDF and perturbative order.

In the future, we plan on improving our framework by implementing also the subleading quark-mass effects. As is the case in inclusive DIS, we can also include radiative electroweak and target-mass corrections in the SIDIS calculation. We are also expecting the full charged-current next-to-NLO corrections to become available soon, as they have recently been made available in the photon-exchange case \cite{Goyal:2023xfi,Bonino:2024qbh,Bonino:2024wgg,Goyal:2024tmo}.

\acknowledgments

We acknowledge the financial support from the Magnus Ehrnrooth foundation (S.Y.), the Research Council of Finland Project No. 331545 (I.H.), and the Center of Excellence in Quark Matter of the Research Council of Finland, project 346326. The reported work is associated with the European Research Council project ERC-2018-ADG-835105 YoctoLHC. We acknowledge grants of computer capacity from the Finnish Grid and Cloud Infrastructure (persistent identifier
   \texttt{urn:nbn:fi:research-infras-2016072533}) and the Finnish IT Center for Science (CSC), under the project jyy2580. 

\bibliographystyle{JHEP}
\bibliography{Refs.bib}

\end{document}